\documentclass[showpacs,floats,aps,twoside,11]{revtex4}
\usepackage{amsmath,amssymb}
\usepackage{fancyhdr,lastpage}
\usepackage{graphicx}
\usepackage{epsfig}
\usepackage{setspace}
\usepackage{float}
\floatstyle{plaintop}
\restylefloat{table}
 
\begin{document}
\setcounter{page}{1}

\title{Discrete variable representation method calculation of the electronic structure of noble gas atoms}

\author{D. Naranchimeg$^1$} 
\author{L. Khenmedekh$^1$} 
\author{G. Munkhsaikhan$^1$}
\author{N. Tsogbadrakh$^2$}
\email{Tsogbadrakh@num.edu.mn}
\affiliation{$^1$Department of physics, Mongolian University of Science and Technology, Ulaanbaatar 14191, Mongolia\\
$^2$Department of physics, National University of Mongolia, Ulaanbaatar 14201, Mongolia}

\begin{abstract}
We have been calculated the ground state charge densities and energies of noble gas atoms through a single time dependent quantum fluid Schr$\ddot{o}$dinger equation. By using imaginary - time, the Schr$\ddot{o}$dinger equation has been transformed into diffusion equation. This equation numerically solved through discrete variable representation (DVR) method. Instead of the usual finite difference method the radial coordinate is discretized using the discrete variable representation constructed from Coulomb wave functions. Our calculations were performed using the Mathematica 7.0 programm.
\end{abstract}

\maketitle           
\thispagestyle{plain}
\setcounter{page}{1}
\goodbreak

\section{Introduction}

Numerical treatment of many - electron systems is extremely computationally demanding task. Density functional theory (DFT) calculation of many - electron systems is opened broad perspective for researchers. Though it uses only three coordinates, the number of equations to solve are increases with the number of electrons to be treated. Instead the quantum fluid density functional theory (QFDFT) solves only one time dependent generalized nonlinear Schr$\ddot{o}$dinger equation (GNLSE) for many - electron systems.
Using imaginary time for GNLSE, one could reach lowest energy state of the system, after tens of iterations.

GNLSE had been solved previously for calculation of ground state properties of noble gas atoms by finite difference method \cite{eConf} and also by generalized pseudospectral method (GPSM) \cite{DPF2009}.
In the present study we used the Coulomb wave function discrete variable representation (CWDVR) method \cite{templates-ref} for solution of the GNLSE for the noble gas atoms.

\section{Theory and Methodology} 

In this section we shall consider the numerical solution of the time - dependent Schr$\ddot{o}$dinger equation for atomic systems to be
\begin{equation}
[-\frac{1}{2}\nabla^2+\upsilon_{eff}(\rho)]\psi(\vec{r,t})=i\frac{\partial\psi(\vec{r,t})}{\partial t}.
\end{equation}
The entire time - evolving interacting system is described by the complex - valued hydrodynamical wave function
\begin{equation}
\psi({\vec{r},t})=\rho(\vec{r},t)^{1/2} e^{i \chi(\vec{r},t)}=R(\vec{r},t) e^{i\chi(\vec{r},t)}.
\end{equation}
However, one can write equation (1) in imaging time $\tau$ and substitute $\tau=-i t$, $t$ being the real time, to obtain, which closely resembles a diffusion - type equation:
\begin{equation}
[-\frac{1}{2}\nabla^2+\upsilon_{eff}(\rho)] R(\vec{r,t})=-\frac{\partial R(\vec{r,t})}{\partial t}
\end{equation}
$R(\vec{r,t})$ is the diffusion function and the diffusion process is governed by $\upsilon_{eff}(\rho)$.\\
$\upsilon_{eff}(\rho)$ contains both classical and quantum potentials
\begin{equation}
\upsilon_{eff}(\rho)=\upsilon_{ee}(\rho)+\upsilon_{ne}(\rho)+\upsilon_{xc}(\rho)+\upsilon_{corr}(\rho)+\upsilon_{ext}(\rho)
\end{equation}
The terms on the right - hand of Eq.(4) are as follows: the first is the inter - electron repulsion term, the second is the electron - nuclear attraction term, the third is exchange - correlation term, fourth term is the nonclassical correction, last term arises from interaction with the external field (in the present case, this interaction is zero).

\section{Numerical Solution: DVR method}

The diffusion equation (3) can be written as
\begin{equation}
[\hat{H_0}(\vec{r})+\hat{V}(\vec{r},t)] R(\vec{r,t})=-\frac{\partial R(\vec{r,t})}{\partial t}.
\end{equation}
We shall extend the second - order split - operator technique in spherical coordinates \cite{arXiv} for the time propagation of the Schr$\ddot{o}$dinger equation:
\begin{equation}
R(\vec{r},t+\Delta t)\simeq e^{-\hat{H_0}\Delta t/2}e^{-\hat{V}(r,t+\Delta t/2)\Delta t}e^{-\hat{H_0}\Delta t/2} R(\vec{r},t)+O(\Delta t^3)
\end{equation}
Here split operator technique is expressed in terms of $\hat{H_0}$, which is chosen to be the radial kinetic operator and $\hat{V}$ the remaining Hamiltonian. Matrix form of Hamiltonian operator is following:
\begin{equation}
[H]_{i j}=(D_2)_{i j}+V(x_{i})\delta_{i j}
\end{equation}
with
\begin{equation}
(D_2)_{i j}=\frac{1}{3}(E+\frac{Z}{x}),   i=j
\end{equation}
\begin{equation}
(D_2)_{i j}=\frac{1}{(x_i-x_j)^2},  i\neq j
\end{equation}
The eigenvalues and eigenfunctions of $\hat{H}$ will be denoted as ${\varepsilon_{k}}$ and ${\phi_{k i}}$, respectively.\\
The propagation of a given radial wave function $R(r,t)$ in order $\hat{H}$ can now be expressed as
\begin{equation}
[e^{-\hat{H}\Delta t/2}R(r,t)]_{i}=\sum_{j=1}^{N}S_{i j}R(r,t)
\end{equation}
where
\begin{equation}
S_{i j}=\sum_{k}\phi_{k i}\phi_{k j} e^{-\varepsilon_{k}\Delta t/2}.
\end{equation}
Note that $S_{i j}$ is a complex symmetric matrix and it needs to be computed only once. The time propagation is therefore reduced to the matrix - vector product, which can be performed efficiently using the Mathematica 7.0 programm.

\section{Results and Discussion}

In this section we present results from nonrelativistic electronic structure calculations of the ground states of noble gas atoms. The main results for He, Ne and Ar atoms are suumarized in the Table I. Results from the calculations for the He atom are well agreed with the results from Roy \cite{arXiv-1} and HF \cite{arXiv-2}. However, in the case of Ne and Ar atoms differences in the total energies are obtained in our calculation, which is the result of the contribution of potential energy calculation. 

Figure 1 represents the result of the calculation of the radial charge density distribution of Ne and Ar atoms. We note that the radial charge density calculated maintain the expected shell structure and closely resemble the HF density (not shown in the plot). 

\begin{center}
\begin{table}
\caption{Calculated ground state properties of He, Ne and Ar(in au) along with literature data for comparison.}
\begin{ruledtabular}
\begin{tabular}{ccccc}
\textbf{}& \textbf{} & \textbf{He} & \textbf{Ne} & \textbf{Ar}\\
\hline -E & Present work & 2.9000 & 128.9990 & 527.1320 \\
 & Roy\cite{arXiv-1} & 2.8973 & 128.9065 & 527.5486 \\
 & HF\cite{arXiv-2} & 2.8617 & 128.5470 & 526.8174 \\
\hline $-Z/r$ & Present work & 6.7878 & 311.115& 1247.2100\\
 & Roy\cite{arXiv-1} & 6.7850 & 311.0597 & 1245.5699 \\
 & HF\cite{arXiv-2} & 6.7492 & 311.1333 & 1255.0504 \\
\hline $1/r_{12}$ & Present work & 2.0678 &65.7672 & 221.4480 \\
 & Roy\cite{arXiv-1} & 2.0651 & 65.7129 & 220.6552 \\
 & HF\cite{arXiv-2} & 2.0516 & 66.1476 & 231.6093 \\
\hline $-E_{x}$ & Present work &1.0273 & 12.1128 &29.5296 \\
 & exact & 1.026 & 12.11 & 30.19 \\
 \hline $-E_c$ & Present work & 0.0422 & 0.3561& 0.7023 \\
 & Roy\cite{arXiv-1} & 0.0423 & 0.3561 & 0.7011  \\
 \hline $T_w$ & Present work & - & 94.2962 & 322.3100 \\
 & Roy\cite{arXiv-1} & - & 94.2068 & 322.0345   \\
 & HF\cite{arXiv-2} & - & 90.6140 & 308.4206  \\
 \hline $T_{corr}$ & Present work & - & 34.7033 & 205.5490 \\
 & Roy\cite{arXiv-1} & - & 34.7006 & 205.5177  \\
 & HF\cite{arXiv-2} & - & 37.3886 & 214.4033  \\
\end{tabular}
\end{ruledtabular}
\label{table1}
\end{table}
\end{center}

\begin{figure}
\includegraphics[width=16.0cm]{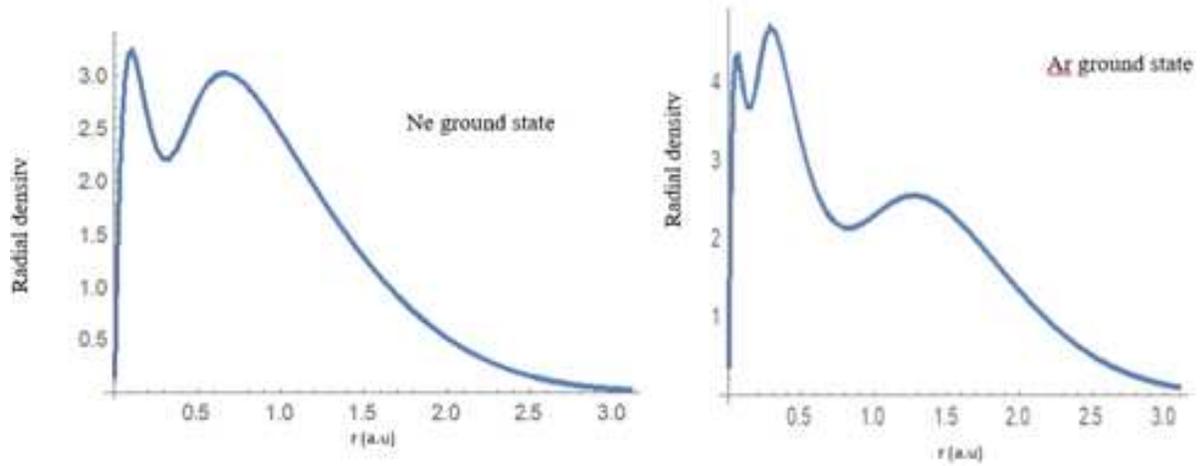}
\caption{The radial charge density distributions of ground state for Neon (left side) and Argon (right side) atoms}
\label{Fig1}
\end{figure}

\section{Conclusions}

The current research is focused on the results of the calculation of the electronic structure of the noble atoms. We describe the discrete variable representation method for the noble gas atoms. Results for noble gas atoms including effective potential, exchange and correlation contributions and ground state energy are presented and compared with calculated results from other researchers. For the next step of the research density functional theory with Kohn-Sham correction for ground state energy of noble gas atoms is going to be calculated.

\begin{acknowledgments}

We have no interested the financial support. 
 
\end{acknowledgments}

\end{document}